\documentstyle[fancyheadings,epsfig,prl,aps,multicol]{revtex}

\makeatletter
\def\@chuckoptarg[#1]{}
\def\figure{
\let\@capwidth\columnwidth
\vskip 1pc
\def\@captype{figure}
\interlinepenalty10000
\@ifnextchar[{\@chuckoptarg}{}
}
\def\endfigure{\goodbreak\vskip1pc}
\@namedef{figure*}{\figure}
\@namedef{endfigure*}{\endfigure}

\def\table{
\let\@capwidth\columnwidth
\vskip 1pc
\def\@captype{table}
\interlinepenalty10000
\@ifnextchar[{\@chuckoptarg}{}
}
\def\endtable{\goodbreak\vskip1pc}
\@namedef{table*}{\table}
\@namedef{endtable*}{\endtable}

\def\widetext{\par\global\columnwidth42.5pc
\global\hsize\columnwidth\global\linewidth\columnwidth
\global\displaywidth\columnwidth}

\widetext

\makeatother

\begin{document}
\draft

\title{Transparency of Magnetized Plasma at Cyclotron Frequency}
\author{G.~Shvets} 
\address{Princeton Plasma Physics Laboratory, Princeton University,
Princeton, NJ 08543}
\author{J.~S.~Wurtele} 
\address{University of California, Berkeley, CA 94720}
\newcommand{\ba}{\begin{eqnarray}}
\newcommand{\ea}{\end{eqnarray}}
\newcommand{\be}{\begin{equation}}
\newcommand{\ee}{\end{equation}}
\newcommand{\para}{\parallel}

\maketitle
\begin{abstract}
Electromagnetic radiation is strongly absorbed by the magnetized 
plasma if its frequency equals the cyclotron frequency of plasma 
electrons. It is demonstrated that absorption can be 
completely canceled in the presence of a second radiation beam, or even 
a magnetostatic field of an undulator, resulting 
in plasma transparency at the cyclotron frequency. 
This effect is reminiscent of the 
electromagnetically-induced transparency (EIT) of the three-level atomic 
systems, except that it occurs in a completely {\it classical} plasma.
Also, because of the complexity of the classical plasma, index of refraction 
at cyclotron frequency differs from unity. 
Potential applications of the EIT in plasma include selective plasma 
heating, electromagnetic control of the index of refraction, and 
electron/ion acceleration.
\end{abstract}

\begin{multicols}{2}
\dimen100=\columnwidth
\setlength{\columnwidth}{3.375in}

Electromagnetically induced transparency (EIT) in quantum-mechanical atomic 
systems is a well understood and thoroughly studied~\cite{eit_orig} 
subject. EIT is the basis of several very important applications, 
such as slow light~\cite{lene_hau_nature}, information transfer between 
matter and light~\cite{fleisch2000,lukin2000}, sound wave generation
~\cite{matsko_sound}, or even testing of the black-hole 
physics~\cite{leonhard_prl}.
Several recent reviews~\cite{eit_review} 
illucidated the quantum mechanical 
mechanism of EIT which relies on the destructive 
interference between several pathways which connect the ground and excited
states of the atom. The purpose of this Letter is to describe EIT in a 
classical plasma.

\begin{figure}[h]
\centering
\epsfig{file=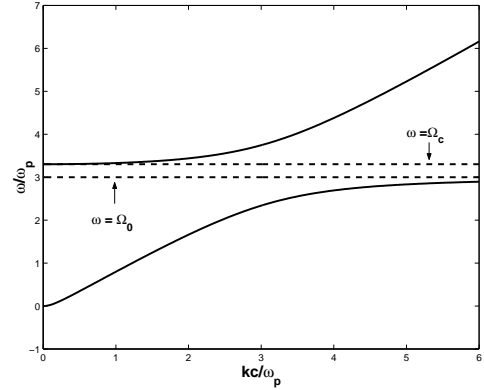, 
angle=-0, width=2.5truein}
\vspace{10pt}
\caption{Dispersion curve for a right-hand polarized wave propagating 
along magnetic field. Forbiden gap exists between cyclotron frequency 
$\Omega_0 = eB_0/mc$ and cutoff frequency 
$\Omega_c = \Omega_0/2 + \sqrt{\Omega_0^2/4 + \omega_p^2}$ }  
\label{fig:dispersion1}
\end{figure}

We consider an externally magnetized plasma with $\vec{B} = B_0 \vec{e}_z$
and density $n_0$. A right-hand polarized electromagnetic wave 
(which we refer to as the probe) at the frequency $\omega_1$ equal 
to cyclotron frequency $\Omega_0 =  eB_0/mc$ cannot propagate in the plasma 
because it undergoes resonant cyclotron absorption~\cite{krall}. 
The cold magnetized 
plasma dispersion relation $\omega_1$ v.~s.$k_1$ for the 
right-hand polarized probe, plotted in 
Fig.~\ref{fig:dispersion1}, is given by
$\displaystyle{
\omega_1^2 = k_1^2c^2  + \frac{\omega_p^2 \omega_1}{\omega_1 - \Omega_0}
}$. Plasma current and the wavenumber $k_1$ become infinite for 
$\omega_1 \rightarrow \Omega_0$, and a forbidden bandgap develops between 
$\Omega_0$ and 
$\Omega_c = \sqrt{\Omega_0^2/4 + \omega_p^2} + \Omega_0/2$, where 
$\omega_p = (4 \pi e^2 n_0/m)^{1/2}$ is the plasma frequency. 
This Letter demonstrates that by adding a second intense 
electromagnetic wave (pump) with frequency 
$\omega_0 = \Omega_0 - \omega_p$ create a transparency near the cyclotron 
frequency. Moreover, if $\omega_p = \Omega_c$, transparency can be created
by a magnetostatic undulator with arbitrary wavenumber $k_0$.

The classical mechanism of the electromagnetically induced 
transparency is the destructive interference between the electric field 
of the probe $\vec{E}_{1 \perp}$ and the sidebands of the electric 
$\vec{E}_{0 \perp}$ and magnetic $\vec{B}_{0 \perp}$ fields of the pump which
are produced by the collective electron plasma oscillation 
with frequency $\omega_p$ along the magnetic field.
Qualitatively, the total force at the cyclotron frequency experienced by 
a plasma electron is given by 
$\vec{F}_{\rm tot} \approx -e(\vec{E}_{1 \perp} + 
\zeta_z \partial_z \vec{E}_{0 \perp} + \dot{\zeta}_z \vec{e}_z \times 
\vec{B}_{0 \perp}$, where $\zeta_z$ is the electron displacement in the 
plasma wave.
If the pump, probe, and plasma waves are properly phased, then 
Therefore, if the amplitudes and phases of the pump and the plasma wave are 
properly correlated, then $\vec{F}_{\rm tot} = 0$. Consequently, 
the plasma current at the cyclotron frequency is small (or even vanishing), 
and the probe propagates as if in vacuum. Our numerical simulation below 
demonstrates that this correlation is naturally achieved in a collisionless
plasma.

We assume two right-hand polarized EM waves propagating along $z-$ direction,
with their electric and magnetic fields given by 
$2e \vec{E}_{0 \perp}/mc\omega_0 = a_{\rm pump} \vec{e}_{+} 
\exp{(i\bar{\theta}_{0})} + c.~c.$, 
$2e \vec{E}_{1 \perp}/mc\omega_1 = a_{\rm probe} \vec{e}_{+} 
\exp{(i\bar{\theta}_{1})} + c.~c.$, and 
$\vec{B}_{0,1 \perp} = (c \vec{k}_{0,1}/\omega_{0,1}) \times \vec{E}_{0,1}$,
where $\vec{e}_{+} = \vec{e}_x + i \vec{e}_y$, 
$\vec{e}_{-} = \vec{e}_x - i \vec{e}_y$, 
$\bar{\theta}_{0} = k_0 z - \omega_0 t$, and 
$\bar{\theta}_{1} = k_1 z - \omega_1 t$. 
Non-relativistic equation of motion of a plasma electron in the combined field is given by
\be
\frac{d^2 \vec{x}}{d t^2} + \Omega_0 \vec{v} \times \vec{e}_z + 
\omega_p^2 \zeta_z \vec{e}_z = -\frac{e}{m} \sum_{m=0,1}\left( 
\vec{E}_m + \frac{\vec{v} \times \vec{B}_{\perp m}}{c} \right), 
\label{eq:exact_motion}
\ee
where $\vec{x} \equiv (z_0 + \zeta_z) \vec{e}_z + \vec{x}_{\perp}$ 
and $\vec{v} = d \vec{x}/dt$ are the particle
position and velocity, and the initial conditions are 
$\vec{v} = 0$ and 
$\vec{x} = z_0 \vec{e}_z$. The third term in the lhs of 
Eq.~(\ref{eq:exact_motion}) is the restoring force of the 
ions~\cite{dawson59}.

Equation (\ref{eq:exact_motion}) was integrated  for two cases:
(a) when only a probe field is turned on, and (b,c) when both the 
pump and the probe are turned on. The pump and the probe amplitudes were 
increased adiabatically in time, up to their respective peak amplitudes of 
$a_0$ and $a_1$, according to 
\ba
&&a_{\rm pump} = \frac{a_0}{2} \left(1 + 
\tanh{[(\Omega_0 t - 160)/40]}\right), \nonumber \\
&&a_{\rm probe} = \frac{a_1}{2} \left( 1 + \tanh{[(\Omega_0 t - 320)/40]}
\right),
\label{eq:rampup}
\ea
enabling the pump to turn on first, followed by the probe.

\begin{figure}[h]
\centering
\epsfig{file=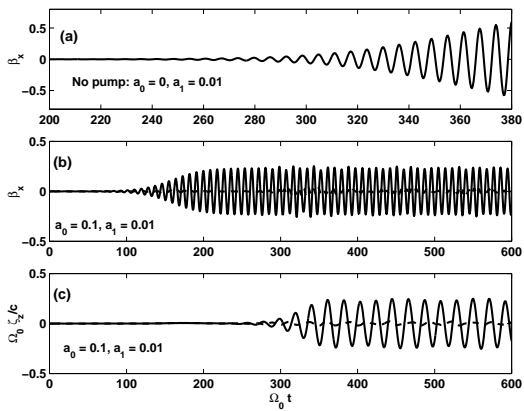, 
angle=-0, width=2.9truein}
\vspace{10pt}
\caption{Numerical simulation of the single particle motion in the combined 
field of two EM waves with ($\omega_1 = \Omega_0$,$k_1 = \omega_1/c$)
and  ($\omega_0 = \Omega_0 - \omega_p$, $k_0 \approx 0.83 \Omega_0/c$). 
Both pump and probe are 
slowly turned on according to Eq.~(\ref{eq:rampup}). 
(a) Without the pump electron is resonantly driven by probe: $\beta_x$ 
growth indefinitely; 
(b) With the pump, electron motion is almost unaffected by the probe. 
Solid line -- total $\beta_x$; 
barely visible dashed line -- $(\beta_x - \beta_{x0})$, where 
$\beta_{x0} = \omega_0 a_{\rm pump}/(\omega_0 - \Omega_0) 
\sin{(k_0 z_0 - \omega_0 t)}$. Since $\beta_x - \beta_{x0}$.
(c) Solid line: longitudinal displacement $\Omega_0 \zeta_z/c$; 
dashed line: $\Omega_0 (\zeta_z - \zeta_0)/c$, where $\zeta_0 = 
2a_{\rm pump}/a_{\rm probe} \sin{\omega_p t}$
from Eq.~(\ref{eq:zeta0_figure})      }  
\label{fig:eit_simulation}
\end{figure}

Simulation results for $\omega_p/\Omega_0 = 0.3$ ($\omega_0 = 0.7 \Omega_0$)
are shown in Fig.~\ref{fig:eit_simulation}. 
Without the pump electron is resonantly driven by the probe as 
shown in Fig.~\ref{fig:eit_simulation}(a). In the plasma, this growth 
manifests itself in a large electron 
current and probe absorption because, time-averaged,
$\vec{E}_{\perp} \cdot \vec{v}_{\perp} < 0$.
Adding a strong pump with $a_0 = 0.1$ and $k_0 \approx 0.83 \Omega_0/c$ 
dramatically changes electron motion, see  Fig.~\ref{fig:eit_simulation}(b). 
After the pump is turned on but before the turning on of the 
probe, electron oscillates in the field of the pump according to 
$\beta_{x0} = \omega_0 a_{\rm pump}/(\omega_0 - \Omega_0) 
\sin{(k_0 z_0 - \omega_0 t)}$. Switching on the probe 
does not significantly alter electron motion: 
$\beta_x - \beta_{x0}$ is shown as a barely visible dashed line 
in Fig.~\ref{fig:eit_simulation}(b)). 
Comparing Figs.~\ref{fig:eit_simulation}(a) and (b), observe
that the pump suppressed electron responce at the cyclotron frequency, making
the plasma transparent to the probe. 

Suppression is caused by the excitation of a strong plasma [shown in 
Fig.~\ref{fig:eit_simulation}(c)] which produced a sideband of the pump 
at the cyclotron frequency. 
This sideband canceled the electric field of the probe. 
An approximate analytic formula for the steady-state amplitude of the 
plasma oscillation,
\be
\zeta_0 = \frac{2a_{\rm probe}}{k_0 a_{\rm pump}} \sin{\omega_p t},
\label{eq:zeta0_figure}
\ee
is derived below by requiring that the sideband cancels the probe. 
which is in good agreement with the simulation result. 
Simulation results demonstrate {\it stability} of the 
steady-state values of $\beta_x$ and $\zeta_z$ which are naturally reached in 
a collisionless plasma.
Note that the pump 
has to be switched on prior to the arrival of the probe. In atomic 
physics, this pulse sequence is referred to as 
``counter-intuitive''~\cite{eit_review}.

\begin{figure}[h]
\centering
\epsfig{file=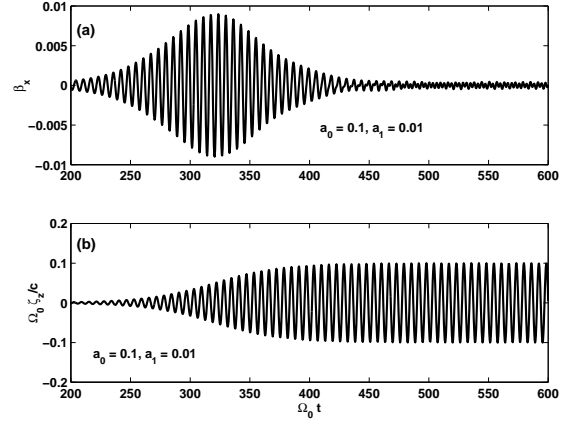, 
angle=-0, width=2.9truein}
\vspace{10pt}
\caption{Same as Fig.~2, except $\omega_p = \Omega_0$,
$\omega_0 = 0$, $k_0 \approx 2 \Omega_0/c$ (static helical undulator is 
switched on from the start). 
(a) Transverse velocity $\beta_x$ and (b) longitudinal displacement 
$\Omega_0 \zeta_z/c$ during and after the turning on of the probe. }
\label{fig:undulator}
\end{figure}

Maintaining high-power pumping waves in the plasma may prove challenging
in practice. For example, supporting $a_0 = 0.01$ over an area 
$A = (2\pi c/\omega_0)^2$ requires microwave power of $3$ megawatts. 
Fortunately, for $\omega_p = \Omega_0$, a magnetostatic helical 
undulator can replace a microwave beam. We simulated 
electron motion in the combined field of an undulator with 
$a_0 = 0.1$ and $k_0 = 2 \Omega_0/c$, and a probe 
which is switched on according to 
$\displaystyle{
a_{\rm probe} = 0.5 a_1 \left( 1 + \tanh{[(\Omega_0 t - 270)/60]}
\right)
}$, where $a_1 = 0.01$. Suppression of the electron 
responce at the cyclotron frequency is apparent from 
Fig.~\ref{fig:undulator}(a). Electric field of the probe is canceled 
by the $(\dot{\zeta}_z/c) \vec{e}_z \times \vec{B}_{0 \perp}$ force 
which is exerted on a longitudinal plasma wave by the helical
magnetic field of the undulator.

Steady-state values of 
$\beta_{+} = \beta_{-}^{\ast} = \beta_x - i \beta_y$ and $\zeta_z$
can be analytically obtained by linearizing Eq.~(\ref{eq:exact_motion}) 
in the weak probe $a_1 \ll a_0$ limit. 
\ba
&&\dot{\beta}_{+} + i \Omega_0 \beta_{+} = \nonumber \\ 
&&- \left( \omega_0 a_0  e^{i\bar{\theta}_0} + 
\omega_1 a_1  e^{i\bar{\theta}_1} - 
k_0 a_0 \dot{\zeta}_z e^{i\bar{\theta}_0} -
k_1 a_1 \dot{\zeta}_z e^{i\bar{\theta}_1}
\right).
\label{eq:vperp2}
\ea

Introducing $\theta_{0,1} = k_{0,1} z_0 - \omega_{0,1} t$ 
and assuming that $k_{0,1} \zeta_z < 1$, exponentials in 
Eq.~(\ref{eq:vperp2}) are expanded as
$\displaystyle{
e^{i\bar{\theta}_{0,1}} \approx e^{i\bar{\theta}_{0,1}} ( 1 + i k_{0,1} 
\zeta_z)}$, yielding
\ba 
&&\dot{\beta}_{+} + i \Omega_0 \beta_{+} = -\omega_0 a_0 e^{i\theta_0} 
\left( 1  + i k_0 \zeta_z - k_0 \dot{\zeta}_z/\omega_0 \right) \nonumber \\ 
&&- \omega_1 a_1 e^{i\theta_1} 
\left( 1  + i k_1 \zeta_z - k_1 \dot{\zeta}_z/\omega_1 \right).
\label{eq:vperp3}
\ea

Longitudinal equation of motion is given by
\[
\ddot{\zeta}_z + \omega_p^2 \zeta_z \approx - \frac{e}{mc} 
\left( \vec{v}_{\perp} \times \vec{B}_{\perp} + \zeta_z \vec{v}_{\perp} \times 
\frac{\partial \vec{B}_{\perp}}{\partial z} \right),
\] 
where $\vec{B}_{\perp}(z,t)$ was expanded as 
$\vec{B}_{\perp}(z_0 + \zeta_z) \approx \vec{B}_{\perp}(z_0) + 
\zeta_z \partial_{z_0} \vec{B}_{\perp}(z_0)$ to first order in 
$\zeta_z$.
Inserting the expression for $\vec{B}_{\perp}$, obtain 
\ba
&&\ddot{\zeta}_z + \omega_p^2 \zeta_z = -\frac{c^2}{2} \left(
k_0 a_0 \beta_{-} e^{i\theta_0} + 
k_1 a_1 \beta_{-} e^{i\theta_1} - \right. \nonumber \\
&&\left. i k_0^2 \zeta_z \beta_{-} a_0 e^{i\theta_0} - 
i k_1^2 \zeta_z a_1 e^{i\theta_1} \right) + c.~c.
\label{eq:zetaz1}
\ea
The last term in the RHS of Eq.~(\ref{eq:zetaz1}) will be later dropped because
it is proportional to the product of two small quantities, $\zeta_z$ and 
$a_1$.
Note that, unlike the transverse velocity $\beta_{+}$ which is excited 
directly by each of the two lasers according to Eq.~(\ref{eq:vperp3}),
plasma waves are excited only in the presense of {\it two} lasers 
via the beatwave mechanisms. 

The physical reason for EIT in plasma is the strong coupling between 
longitudinal and transverse degrees of freedom of the plasma electrons.
The steady-state solution of Eq.~(\ref{eq:zetaz1})  
$\displaystyle{\zeta_z = 
0.5 \tilde{\zeta} \exp{i(\Delta k z - \Delta \omega t)} + c.~c.~}$, 
where $\Delta \omega = \omega_1 - \omega_0$ and 
$\Delta k = k_1 - k_0$, is substituted into 
the transverse equation 
of motion (\ref{eq:vperp3}). Retaining the terms with 
$\exp{-i\omega_0 t}$ and $\exp{-i\omega_1 t}$ dependence results in
\be
\beta_{+} = - \frac{i \omega_0 a_0}{\omega_0 - \Omega_0} 
e^{i \theta_0} - 
\frac{i \omega_1}{\omega_1 - \Omega_0} \left( a_1 + 
\frac{ik_0 \tilde{\zeta}}{2} a_0 \right) e^{i \theta_1}. 
\label{eq:betaplus2}
\ee
Applying Eq.~(\ref{eq:betaplus2}) to the simulated earlier case of 
$\omega_1 = \Omega_0$ and $\Delta \omega = \omega_p$ yields the steady-state 
amplitude of the plasma wave given by Eq.~(\ref{eq:zeta0_figure}).

In the general case of $\omega_1 \neq \Omega_0$ we 
insert $\beta_{+}$ and $\beta_{-}$ into Eq.~(\ref{eq:zetaz1}) yielding
\ba
&&(\omega_p^2 - \Delta \omega^2) \tilde{\zeta} = i c^2 
\left[ \frac{k_0 a_0^{\ast} \omega_1}{\omega_1 - \Omega_0} 
(a_1 + i k_0 \tilde{\zeta} a_0/2) - \right. \nonumber \\
&&\left. \frac{k_1 a_1 \omega_0}{\omega_0 - \Omega_0} 
a_0^{\ast} - i \frac{k_0^2 \tilde{\zeta} \omega_0}{\omega_0 - \Omega_0}
|a_0|^2 \right], 
\label{eq:zetaz2} 
\ea
where $\theta_1 - \theta_0 = (k_1 - k_0) z_0 - \Delta\omega t$. 
Equation (\ref{eq:zetaz2}) is then solved for $\tilde{\zeta}$ which is 
substituted into Eq.~(\ref{eq:betaplus2}) yielding the steady-state value of 
$\beta_{+}$:
\ba
&&\beta_{+s} = - \frac{i \omega_0 a_0}{\omega_0 - \Omega_0} e^{i \theta_0} 
 - i \frac{\omega_1 a_1}{\omega_0 - \Omega_0} e^{i \theta_1} \times
\nonumber \\ 
&& \times
\frac{c^2 k_0^2 \omega_0 |a_0|^2 (k_1/k_0 - 2) + 
2(\omega_p^2 - \Delta\omega^2)(\omega_0 - \Omega_0)}{
c^2 k_0^2 |a_0|^2 \omega_1 +
2(\omega_p^2 - \Delta\omega^2)(\omega_1 - \Omega_0) },
\label{eq:betas1}
\ea
%
%\be
%\tilde{\zeta}_s = 2 i c^2 a_0^{\ast} a_1 \frac{
%k_0 \omega_1 - k_1 \omega_0 (\omega_1 - \Omega_0)/(\omega_0-\Omega_0) }{2
%(\omega_p^2 - \Delta\omega^2)(\omega_1 - \Omega_0) + c^2 k_0^2 
%|a_0|^2 \omega_1 },
%\label{eq:zetaz_stable2}
%\ee
where we have neglected terms proportional to the product of laser 
detuning $\delta \Omega = \omega_1 - \Omega_0$ from resonance and the pump 
intensity $a_0^2$. Qualitatively, the pump influence is 
strong only close to the 
cyclotron resonance, and is negligible far from $\omega_1 = \Omega_0$.
From Eq.~(\ref{eq:betas1}), plasma is resonantly driven 
when the denominator 
$D = 2(\omega_p^2 - \Delta\omega^2)(\omega_1 - \Omega_0) + c^2 k_0^2 
|a_0|^2 \omega_1$ vanishes. Close to cyclotron resonance  
$D \approx 4 \omega_p (\Omega_R^2 - \delta \Omega^2)$, where 
$\Omega_R = c k_0 a_0 (\Omega_0/4\omega_p)^{1/2}$ is the effective 
Rabi frequency. Hence, the modified plasma resonances are shifted 
from $\omega_1 = \Omega_0$ to $\omega_{1} = \Omega_0 \pm \Omega_R$. 
Fluid velocity component $b_{+} \approx \beta_{+s} - 
\partial_z (\zeta_z \beta_{+s}) = \beta_{+}-i k_1\zeta_z \beta_{+}$ 
proportional to $\exp{i\theta_1}$  is given by  
$\displaystyle{
b_{+} = i a_1 \omega_1 
(\delta \Omega + \delta \Omega_0)/(\Omega_R^2 - \delta \Omega^2)
}$
where $\delta \Omega_0(k_1) = 
(2 \Omega_R^2 \omega_0/\omega_p \Omega_0)(k_1/k_0 - 1)$. 

Dispersion relation for classical EIT in magnetized plasma 
is derived from the 
wave equation for the probe 
$\displaystyle{
- (c^2 \partial_z^2 - \partial_t^2) \vec{E} = 4 \pi \partial_t \vec{J}
}$, where the rhs is equal to $-4\pi i c \omega_1 e n_0 b_{+} + 
c.~c.$:
\be
\omega_1^2 = c^2 k_1^2 - \omega_p^2 \omega_1 \
\frac{\delta\Omega + 
\delta \Omega_0(k_1)}{\Omega_R^2 - (\delta\Omega)^2},
\label{eq:disp11}
\ee
where it was assumed that the frequency of the pump is fixed at 
$\omega_0 = \Omega_0 - \omega_p$.
Complete transparency ($\omega_1 = k_1^2 c^2$) is achieved at 
$\omega_1 = \Omega_0 - \delta \bar{\Omega}_0$, where 
$\delta \bar{\Omega}_0 \approx (2 \omega_0 \Omega_R^2/\omega_p \Omega_0)
(\Omega_0/k_0c - 1)$. Note that this frequency shift is in general very small 
in the most interesting regime of $\Omega_R \ll \omega_p$: 
$|\delta \Omega_0| < 4 \Omega_R^2/\omega_p \ll \Omega_R$, and can be even 
smaller near cyclotron resonance when pump and probe co-propagate.  
Equation (\ref{eq:disp11}) reduces to the dispersion relation for a single 
probe in magnetized plasma for
large detunings $(\delta \Omega)^2 \gg \Omega_R^2$. The 
influence of the pump is significant only in the vicinity of
$\delta\Omega = 0$. 

Note that index of refraction is not identically equal to 
unity at the cyclotron resonance. 
This is different from the quantum-mechanical 
result for a three-level system~\cite{scully}, where $\omega_1 = k_1 c$ on 
resonance.  
It can be demonstrated that this difference occurs because 
multiple Landau levels $E_n = n \hbar \Omega_0$ and corresponding 
Raman-shifted levels $E_{Rn} = E_n + \hbar \omega_p$ 
participate in the classical EIT. 

\begin{figure}[h]
\centering
\epsfig{file=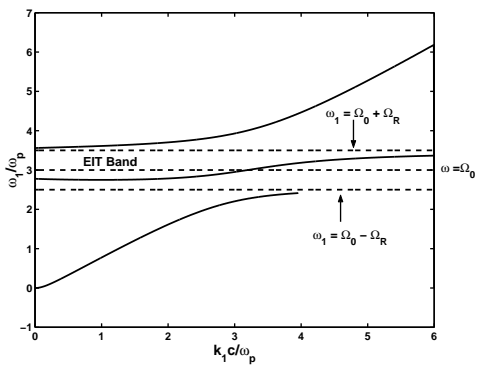, 
angle=-0, width=2.5truein}
\vspace{10pt}
\caption{EIT dispersion curve, $\Omega_0/\omega_p = 3$ and 
$\Omega_R/\omega_p = 1/2$. 
Flat band above $\Omega_0 - \Omega_R$ up to 
$\Omega_0 + \Omega_R$ labeled ``EIT Band'' corresponds to ``slow light'' and
appears only in the presence of a pump. }  
\label{fig:eit_dispersion}
\end{figure}

Dispersion relation given by Eq.~(\ref{eq:disp11}) is plotted in 
Fig.~(\ref{fig:eit_dispersion}) for the same plasma parameters as 
in Fig.~(\ref{fig:dispersion1}), plus a co-propagating pump 
with $\Omega_R = 0.5 \omega_p$. The flat band between the 
$\Omega_0 \pm \Omega_R$ resonant frequencies is a novel feature which is 
not present without the pump (compare with Fig.~(\ref{fig:dispersion1})).
The width of this EIT band proportional to $\Omega_R \propto a_0$ can 
become very narrow for low pump amplitude. The corresponding ``group 
velocity'' (understood in a stricly geometrical sense explained below) 
$v_g = \partial \omega_1/\partial k_1 \approx 2 c \Omega_R^2/\omega_p^2$ 
can also be made arbitrarily small.  Slowly propagating wavepacket of 
electromagnetic waves is a classical analog of the ``slow light'' in 
atomic systems~\cite{lene_hau_nature}.

Qualitatively, the spectacular slowing down of EM waves in the EIT plasma
can be understood by considering the entrance of a probe beam of 
duration $L_0$ into the plasma. 
In steady state inside the plasma, the ``slow light'' wavepacket of 
length $L_f$ consists of the transversely polarized field of the probe 
$|\vec{E}_1| = |\vec{B}_1| = a_1 mc \omega_1/e$ and the longitudinal 
electric field of the plasma wave $E_z = 4\pi e n_0 (2a_1/k_0 a_0)$.
As the pulse enters the plasma, it loses photons to the pump at the 
same rate as new plasmons are created (according to the Manley-Rowe 
relation). Classical photon density of a field with frequency 
$\omega$ is proportional to the action density
$\propto U/\omega$, where $U$ is the energy density. We calculate that the 
ratio of the plasmon to photon density inside the ``slow light'' pulse,
\be
\frac{U_{\rm plas}/\omega_p}{U_{\rm phot}/\omega_1} = 
\frac{\Omega_0}{\omega_p} \frac{E_z^2}{2 E_1^2} = 
\frac{\omega_p^2}{2\Omega_R^2},
\label{eq:photon_account1}
\ee
is $\gg 1$ if $\Omega_R \ll \omega_p$. Thus, most photons of the original 
pulse are lost to the pump. Since the index of refraction remains close 
to unity, so is the photon energy density. Therefore, the loss of photons
is due to the spatial shortening of the pulse from $L_0$ to 
$L_f = L_0 \times (2\Omega_R^2/\omega_p^2)$. Because temporal pulse duration
does not change, we recover the previously calculated 
$v_g/c = 2\Omega_R^2/\omega_p^2$. It is precisely in this 
geometric sense of $v_g/c = L_f/L_0$ that the group velocity of the slow light
is interpreted. $v_g$ is not related to the speed of individual photons 
since their number is not conserved during the pulse transition into the 
plasma.

One interesting application of EIT in magnetized plasma is ion acceleration.
While laser-plasma accelerators of electrons~\cite{tajima} 
have long been considered as a long-term alternative to conventional
rf cavity-based linacs, the field of plasma-based ion accelerators is still 
in its infancy~\cite{cowan}. EIT enables one to conceive a short-pulse 
ion accelerator which consists of a ``slow light'' pulse in plasma 
with approximately equal group and phase velocities. Acceleration is 
accomplished by the longitudinal electric field of the plasma wave. 
Counter-propagating geometry is chosen to match the phase and group 
velocities because 
$v_{\rm ph} = \omega_p/|k_0|+k_1 \approx 0.5 c \omega_p/\Omega_0$. 
Matching $v_{\rm ph} = v_g$ yields $a_0 \approx \omega_p^2/\Omega_0^2 \ll 1$.
Other types of accelerators based on the ``slow light'' 
which rely on the ponderomotive force also
appear attractive because the ponderomotive force, which
scales as the gradient of the energy density 
$E_z^2/L_f \propto (\omega_p/\Omega_0) U_0/v_g^2$, increases rapidly 
with decreasing group velocity of the probe.

\end{multicols}
\end{document}